\documentstyle[times,12pt,epsfig]{article}

\begin{document}

\title{Similarity of percolation thresholds on the hcp and fcc lattices}
\author{Christian D. Lorenz \thanks{cdl@umich.edu}, Raechelle May, and Robert M.
Ziff \thanks{rziff@umich.edu}\\
{\small\it Department of Chemical Engineering, University of Michigan}\\
{\small\it Ann Arbor, MI}}

\date{\today}
\maketitle

\begin{abstract}
Extensive Monte-Carlo simulations were performed in order to determine the precise values of the
critical thresholds for site ($p^{hcp}_{c,S} = 0.199\,255\,5 \pm 0.000\,001\,0$) and bond
($p^{hcp}_{c,B} = 0.120\,164\,0 \pm 0.000\,001\,0$) percolation on the hcp
lattice to compare with previous precise measuremens on the fcc lattice.  Also, exact enumeration of
the hcp and fcc lattices was performed and yielded generating functions and series for the
zeroth, first, and second moments of both lattices.  When these series and the values of $p_c$ are
compared to those for the fcc lattice, it is apparent that the site percolation thresholds are
different; however, the bond percolation thresholds are equal within error bars, and the series
only differ slightly in the higher order terms, suggesting the actual values are very close to each
other, if not identical. 
\end{abstract}

\section{Introduction}
\label{sec1}

The percolation model is used to describe many problems that include a connectivity probabilty,
particularly flow through porous media \cite{SA,Sahimi}.  The three-dimensional lattices that are
often used to model porous media, include simple cubic, body-centered cubic,
face-centered cubic (fcc) and hexagonal close-packed (hcp).  

In a recent paper, Tarasevich and van der Marck \cite{vanderMarck699} pointed out that the critical
thresholds for site and bond percolation on the hcp
lattice were equal within the known error bars to the thresholds for the fcc lattice.  However, the
values for the hcp lattice ($p^{hcp}_{c,S} = 0.1199\,24 \pm 0.000\,05$
\cite{vanderMarck97}, $p^{hcp}_{c,B} = 0.120\,15 \pm 0.000\,05$ \cite{vanderMarck97}) are not as
precise as those found for the fcc lattice ($p^{fcc}_{c,S} = 0.199\,236\,5 \pm 0.000\,001\,0$
\cite{cdlrzjpa},
$p^{fcc}_{c,B} = 0.120\,163\,5 \pm 0.000\,001\,0$ \cite{cdlrzpre}). This raises the very interesting
question of whether or not the thresholds of these two lattices might in fact be identical, or at
least the same to a very high precision.

\begin{figure}
\centerline{\epsfig{file=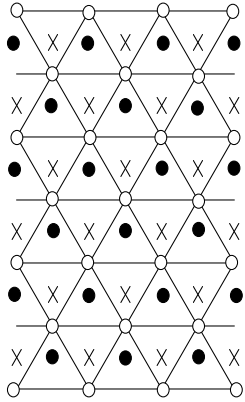, height=150pt}}
\caption{Comparison of the layers that form the hcp and fcc structures.  Both of the structures have the
empty circles (A) as one layer.  However, the hcp lattice has a layer of the darkened circles (B)
above and below that layer; whereas, the fcc lattice has a layer of the darkened circles (B) above
and a layer of the X's (C) below.}
\label{compare}
\end{figure}

The similarity of the critical thresholds for the hcp and fcc lattices  could be explained by the
relative similarity of the hcp and fcc structures.   In Figure \ref{compare}, the hcp and fcc
structures are shown in relation to one another.  The two structures share the same first two layers
(labeled $A$ and $B$ in Fig.\ \ref{compare}).  The difference between the two structures occurs in
the third layer.  The third layer of the fcc is a unique layer which fills the holes in the first
layer which were not filled by the second layer (labeled $C$ in Fig.\ \ref{compare}), but the third
layer of the hcp is the same as layer $A$ \cite{kittel}.  The structure of the hcp crystal can then
be summarized as $A,B,A,B,A,B,\ldots$ and the fcc crystal is
$A,B,C,A,B,C,\ldots$.  Therefore, we pondered whether it took higher precision to see this
strucutral difference or if that difference had no impact on the percolation thresholds of the
two lattices.  
 
In order to extend this study to higher precision we used a growth or epidemic analysis to determine
precise values of the critical thresholds of the hcp lattice, which were then compared to those of
the fcc lattice.  We also carried out an exact enumeration study, because we couldn't find any such
series expansions for the hcp and fcc lattices in literature; we could only find an exact
enumeration study of a modified fcc lattice \cite{bocquet}.

In the following sections, we report on the determination of the new values of $p_c$ for the hcp
lattice and the details of the exact enumeration.  The
results are summarized and discussed in the conclusion section.

\section{Percolation thresholds}
\label{sec2}

Precise values of the thresholds for bond and site percolation on the hcp lattice were found
using procedures similar to those outlined  for site percolation in \cite{cdlrzjpa} and for bond
percolation in
\cite{cdlrzpre}.   A virtual lattice of
$2048^{3}$ sites was simulated, using the block-data method first described in
\cite{ZCS}. We distorted both lattices so that all sites fell on a
simple cubic lattice.  On these lattices, we grew individual clusters by a Leath-type algorithm which
used the unit vectors shown in Table
\ref{unitvectors} for the hcp and fcc lattices.  For the hcp lattice, the unit vectors in the plane
will always be the same, but the unit vectors needed to check the nearest neighbors above or below a
certain site depends on which level the site is located ($A$ or $B$).  In Table \ref{unitvectors},
the unit vectors required to check the nearest neighbors in the plane and when going from layer $A$
to $B$ and from $B$ to $A$ are shown for the hcp lattice.  The critical thresholds were identified
using an epidemic scaling analysis.  In order to determine the critical thresholds at the reported
precision,
 about $2 \times 10^{7}$ clusters were generated utilizing about $10^{13}$ random numbers,
which required a few weeks worth of computer time on ten workstations. 

\begin{table}
\caption{Unit vectors used to describe the neighbors in the fcc and hcp lattices.
\label{unitvectors}}
\centerline{\begin{tabular}{|ll|} \hline \hline
Lattice&Vectors\\ \hline
fcc&$(1,1,0)$, $(1,-1,0)$, $(-1,-1,0)$, $(-1,1,0)$, $(1,0,1)$, $(-1,0,1)$, $(1,0,-1)$, \\
&$(-1,0,-1)$,$(0,1,1)$, $(0,-1,1)$, $(0,1,-1)$, $(0,-1,-1)$\\ \hline
hcp (in the plane)&$(1,0,0)$, $(1,1,0)$, $(0,1,0)$, $(-1,0,0)$, $(-1,-1,0)$, $(0,-1,0)$, \\
&\\
(A to B)&$(0,0,1)$, $(0,0,-1)$, $(0,-1,1)$, $(0,-1,-1)$, $(1,0,1)$, $(1,0,-1)$\\
&\\ 
(B to A)&$(0,0,1)$, $(0,0,-1)$, $(-1,0,1)$, $(-1,0,-1)$, $(0,1,1)$, $(0,1,-1)$\\ 
\hline \hline
\end{tabular}}
\end{table}

\begin{figure}
\centerline{\epsfig{file=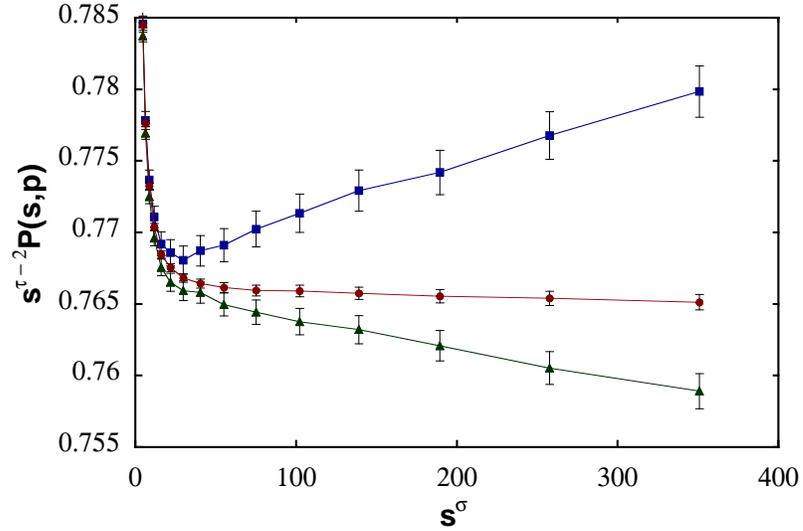, height=200pt}}
\caption{Plot of $s^{\tau-2}P(s,p)$ versus $s^\sigma$ for bond percolation on the hcp
lattice.  The curves plotted here represent $p=0.120\,170\,0$, $p=0.120\,163\,5$, and
$p=0.120\,160\,0$ (from top to bottom) respectively.}
\label{bondpc}
\end{figure}
\begin{figure}
\centerline{\epsfig{file=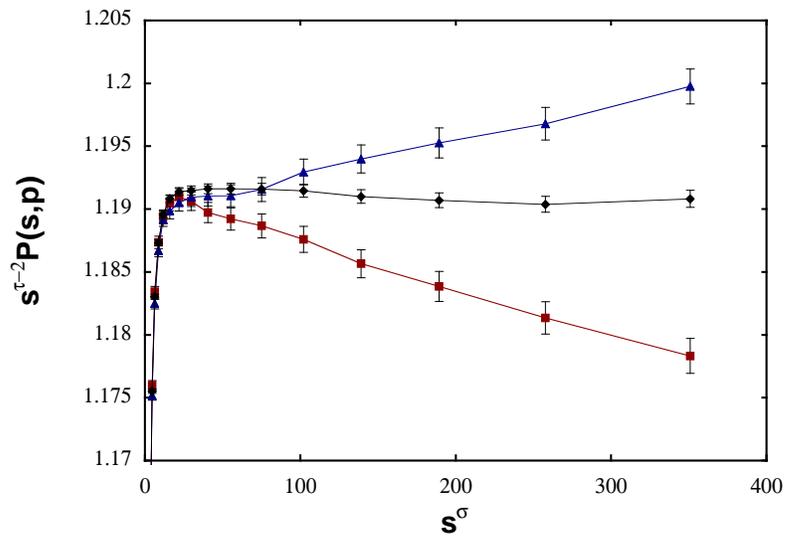, height=200pt}}
\caption{Plot of $s^{\tau-2}P(s,p)$ versus $s^\sigma$ for site percolation on the hcp
lattice.  The curves plotted here represent $p=0.199\,260\,0$, $p=0.199\,255\,5$, and
$p=0.199\,250\,0$ (from top to bottom) respectively.}
\label{sitepc}
\end{figure}

The simulation was used to find the fraction of clusters $P(s,p)$ that grew to a size greater than or
equal to
$s$ sites.  When $p$ is near $p_c$, one expects $P(s,p)$ to behave as 
\begin{equation}
P(s,p) \sim As^{2-\tau}f((p-p_{c})s^{\sigma}) \approx As^{2-\tau}[1 + C(p-p_{c})s^{\sigma} + \ldots]
\label{scaling}
\end{equation}
where $\tau$ and $\sigma$ are universal exponents \cite{SA,fisher}.  We assumed the values $\tau =
2.189$ and $\sigma = 0.445$, which are consistent with other three-dimensional percolation studies
\cite{cdlrzjpa,cdlrzpre,rzgs,ballesteros}. Plots of $s^{\tau - 2}P(s,p)$ versus $s^{\sigma}$ for site
and bond percolation on the hcp lattice were used to find the value of the percolation threshold which
corresponds to horizontal behavior for large $s$.  The results are plotted in Figures
\ref{bondpc} and \ref{sitepc} and imply the following values for the critical thresholds for site
(S) and bond (B) percolation:
\begin{eqnarray}
		&&p^{hcp}_{c,S} = 0.199\,255\,5 \pm 0.000\,001\,0 \nonumber \\
		&&p^{hcp}_{c,B} = 0.120\,164\,0 \pm 0.000\,001\,0.
\label{criticalthresholds}
\end{eqnarray}
For the fcc lattice, we previously found the values \cite{cdlrzjpa,cdlrzpre}:
\begin{eqnarray}
		&&p^{fcc}_{c,S} = 0.199\,236\,5 \pm 0.000\,001\,0 \nonumber \\
		&&p^{fcc}_{c,B} = 0.120\,163\,5 \pm 0.000\,001\,0.
\label{fccthresholds}
\end{eqnarray}
The site thresholds for these two lattices differ by only $0.000\,019$, which is
statistically significant being nearly 10 combined error bars apart.  The bond thresholds, on the
other hand, are identical within the error bars.

\section{Exact enumeration studies of the hcp and fcc lattices}
\label{sec3}

The similarity of the thresholds for the hcp and fcc lattices 
led us to also carry out an exact enumeration calculation, to see how the series of the two lattices
compare.  In exact enumeration, the problem is to find
$g_{st}$, the number of clusters containing $s$ occupied sites and $t$ vacant 
neighboring sites or bonds.  Knowing $g_{st}$, one can find the number of clusters (per site) containing
$s$ occupied sites by
\begin{equation}
 n_s(p) = \sum_t g_{st}p^s q^t\ , 
\label{exact1}
\end{equation}
the total number of clusters per site,
\begin{equation}
N = M_0(p) =\sum_{s} n_{s} = \sum_{s,t} g_{st}p^s q^t\ , 
\label{exact2}
\end{equation}
the percolation probability, 
\begin{equation}
 P(q) = p - M_1(p) = p - \sum_{s} s n_{s} = p - \sum_{s,t} s g_{st}p^s q^t\ ,
\label{exact3}
\end{equation}
(for small $q$), and the susceptibility,
\begin{equation}
 S(p) = M_2(p) = \sum_{s} s^2 n_{s} = \sum_{s,t} s^2 g_{st}p^s q^t\ ,
\label{exact4}
\end{equation}
where $q = 1 - p$.  To calculate all these quantities, it is useful to construct
the generating function 
\begin{equation}
G(p,q) = \sum_{s,t} g_{st}p^s q^t\ .
\label{exact5}
\end{equation}

To find $g_{st}$,  we developed a rather simple
enumeration method based upon the cluster-growth
algorithm, but using a deterministic sequence to decide whether each
successive site (or bond) is occupied or vacant.  After a cluster was 
finished, we stepped back to the last vacant site and made
it occupied (if $s$ was below the cutoff), or stepped back
through the last group of occupied sites to first vacant one
before it, and made that site occupied (if $s$ was at the cutoff),
and then returned to the growth algorithm again.
In this way we went through a binary search of all possible
growth scenarios.  This method is similar to the algorithm described by Redner \cite{redner}.  We
tested our algorithm with published results \cite{mertens} in 2- and 3-dimensions, and found
agreement.  Although  slower than Merten's Fortran code \cite{mertens}, our program was easy
to code and generalize for the different
lattices and both site and bond percolation.  
We found the various series to $s=9$ in a few days of computer time.

\begin{figure}
\centerline{\epsfig{file=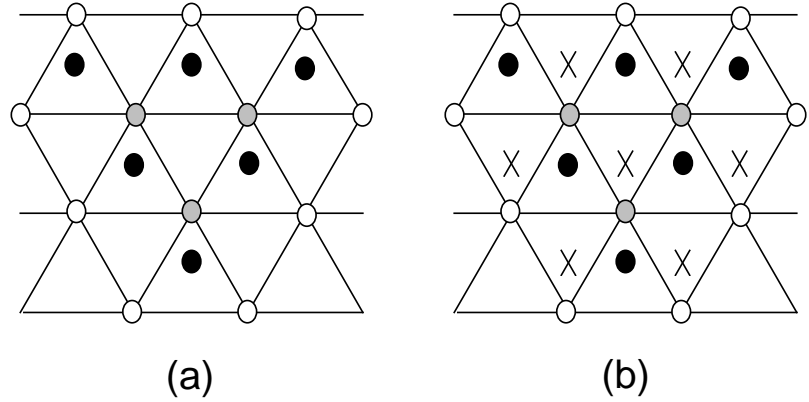, height=150pt}}
\caption{A triangular cluster of three sites (gray colored circles) on the
(a) hcp and (b) fcc lattice.  In (a), the empty circles represent vacant sites in the
plane of the cluster, the black circles represent vacant sites above and below the plane. 
Neighboring this cluster, there are 9 perimeter sites in the plane, 6 perimeter sites in the plane
above and 6 in the plane below, yielding $t = 9+6+6 = 21$ total perimeter sites.  In (b), the black
circles represent the sites above the plane, and the X's represent the sites below the plane.  This
lattice has 9 perimeter sites in the plane, 6 in the plane above, and  7 in the plane below,
yielding $t = 9+6+7 = 22$ total perimeter sites.}
\label{nn}
\end{figure}

For bond percolation, we consider that $s$ represents the number of
occupied sites in a cluster, irrespective of the number of bonds
that are needed to connect them, and $t$ represents
the number of vacant bonds. We thus define $g_{st}$ as the number of clusters
containing $s$ occupied sites and $t$ vacant perimeter bonds.  Note that no bond
is placed in internal, redundant locations.  A
cluster  containing $s$ occupied sites and $t$ vacant perimeter bonds
has a weight $g_{st}p^{s-1}q^t$, because only $s-1$
occupied bonds are needed to connect the $s$ sites.  Therefore,
in bond percolation, all definitions (\ref{exact1})-(\ref{exact4})  should
actually be  divided by $p$ on the right-hand-side, or equivalently those definitions actually give
$p M_n(p)$.

First consider the $g_{st}$ itself, 
which we represent by the generating function (\ref{exact5}), where the coefficients of $p^{s}$ are
the perimeter polynomials. Up to order $s = 5$, the results are:

\bigskip
\noindent$G^{fcc}_{S} = p{q}^{12}+6{p}^{2} {q}^{18}+{p}^{3} (8{q}^{22}+12{q}^{23}+30
{q}^{24}) +{p}^{4}( 2{q}^{24}+27{q}^
{26}+48{q}^{27}+96{q}^{28}+144{q}^{29}+158{q}^{30})$
$+{p}^{5} (24{q}^{28}+6{q}^{29}+132{q}^{30}+264{q}^{31}+423{q}^{32}$
$+780{q}^{33}+1194{q}^{34}+1212{q}^{35}+846 {q}^{36}) +\ldots $

\bigskip
\noindent$G^{hcp}_{S}
=p{q}^{12}+6{p}^{2}{q}^{18}+{p}^{3}({q}^{21}+6{q}^{22}+
13{q}^{23}+30{q}^{24})+{p}^{4}( 8{q}^{25}
+21{q}^{26}+48{q}^{27}+90{q}^{28}+
150 {q}^{29}+158{q}^{30})$
$+{p}^{5}({q}^{27}+12{q}^{28}+36{q}^{29}+114{q}^{30}+303{q}^{31}+357{q}^{32}+801{q}^{33}+1140{q}^{34}+1278{q}
^{35}+840{q}^{36}) +\ldots$

\bigskip
\noindent$G^{fcc}_{B}= p{q}^{12}+6{p}^{2}{q}^{22} +{p}^{3} (8{q}^{30}+16{q}^{31}+42{q}^{32})$
$+{p}^{4}(2{q}^{36}+6{q}^{37}+36{q}^{38}+84{q}^{39}+219{q}^{40}+249{q}^{41}+326{q}^{42})$
$+{p}^{5}(30{q}^{44}+120{q}^{45}+372{q}^{46}+792
{q}^{47}+1596{q}^{48}+2328{q}^{49}+3576{q}^{50}+3072{q}^{51}+2739{q}^{52}) +\ldots$

\bigskip
\noindent$G^{hcp}_{B}=pq^{12} + 6p^2{q}^{22}+
p^3(8{q}^{30}+16{q}^{31}+42{q}^{32})$
$+p^4(2{q}^{36}+6{q}^{37}+36{q}^{38}+84{q}^{39}+219{q}^{40}+
249{q}^{41}+326{q}^{42})$
$+p^5({q}^{42}+4{q}^{43}+34{q}^{44}+114{q}^{45}+357{q}^{46}+780{q}^{47}+1611{q}^{48}+2382{q}^{49}+3513{q}
^{50}+3090{q}^{51}+2739{q}^{52}) +\ldots$
\bigskip

For site percolation, the differences between the two lattices start to show up with $s=3$.
For example, the term
for $s=3$ and $t=21$ occurs for the hcp but not the fcc lattice.
This corresponds to the cluster of three sites in a triangle on the plane,
as shown in Fig. \ref{nn}.\  For bond percolation, the first difference does not occur 
until $s=5$.

First we compare the moments for site percolation, writing each moment
for the two different lattices adjacent to each other for easy comparison:

$$M_{0,S}^{fcc}=p-6p^{2}+8p^{3}+p^{4}-6p^{5}+30p^{6}-4p^{7}+105p^{8}+79p
^{9}\dots$$

$$M_{0,S}^{hcp}=p-6p^{2}+8p^{3}+p^{4}-6p^{5}+30p^{6}+2p^{7}+33p^{8}+513p
^{9}\dots$$

$$M_{1,S}^{fcc}/p=q^{12}+12q^{18}-12q^{19}+24q^{22}-12q^{23}+50q^{24}-168q^{
25}+222q^{26}-140q^{27}+252q^{28}+\dots$$

$$M_{1,S}^{hcp}/p=q^{12}+12q^{18}-12q^{19}+3q^{21}+12q^{22}+6q^{23}+30q^{24}
-109q^{25}+78q^{26}+41q^{27}+44q^{28}+\dots$$

$$M_{2,S}^{fcc} = p+12p^{2}+84p^{3}+504p^{4}+3012p^{5}+17142p^{6}+96228p^{7}
+532028p^{8}+2918388p^{9}+\ldots$$

$$M_{2,S}^{hcp} = p+12p^{2}+84p^{3}+504p^{4}+3014p^{5}+17148p^{6}+96072p^{7}
+533286p^{8}+2911166p^{9}+\ldots$$

For the first moment, we report $M_1/p$, which gives somewhat simpler
expressions for the $q$ series than $M_1$.  Note that the series are
identical up to order $p^6$ for $M_0$, $q^{19}$ (three terms)
for $M_1/p$, and $p^4$ for $M_2$.  The coefficients for  $M_2$
differ a small amount between the two lattices for higher 
order.  Note that the series for $M_1(q)$ is actually given up to order
38 by the enumerations for $s \le 9$.

For bond percolation, the corresponding series are:
$$pM_{0,B}^{fcc} = p-6{p}^{2}+8{p}^{4}+33{p}^{5}+132{p}^{6}+554{p}^{7}
+2514{p}^{8}+13152{p}^{9} +\ldots$$
$$pM_{0,B}^{hcp} = p-6p^{2}+8p^{4}+33p^{5}+132p^{6}+553p^{7}
+2526p^{8}+13116p^{9} + \dots $$

\bigskip
 $ M_{1,B}^{fcc} =
{q}^{12}+12{q}^{22}-12{q}^{23}+24{q}^{30}+54{q}^{32}-204{q}^{33}+126{q}^{34}+8{q}^{36}+96{q}^{38}-32{q}
^{39}+276{q}^{40}-768{q}^{41}+608{q}^{42}-1800{q}^{43}+3066{q}^{44}-1304{q}^{45}+360{q}^{46}-480{q}^{47}+
1056{q}^{48}-3360{q}^{49}+\ldots$

\bigskip
\noindent $ M_{1,B}^{hcp} = q^{12}+12q^{22}-12q^{23}+24q^{30}+54q^{32}-204q^{33}+126q^
{34}+8q^{36}+96q^{38}-32q^{39}+276q^{40}-768q^{41}+613q^{
42}-1800q^{43}+3036q^{44}-1314q^{45}+450q^{46}-480q^{47}+
1061q^{48}-3480q^{49}+\ldots$
$$p M_{2,B}^{fcc}=p+12{p}^{2}+132{p}^{3}+1356{p}^{4}+13344{p}^{5}+127548
{p}^{6}+1194864{p}^{7}+11033256{p}^{8}+100692522{p}^{9}+\ldots$$
$$p M_{2,B}^{hcp} = p+12p^{2}+132p^{3}+1356p^{4}+13344p^{5}+127548p^{6}+1194944
p^{7}+11033544p^{8}+100697070p^{9}+\ldots$$

Here we find an even closer agreement between the series of the two lattices than we found for site
percolation. The series for both $pM_0$ and $pM_2$ agree between the two
lattices up to order 6 and then differ by a very
small amount for the three orders beyond that, while the series for $M_1$
agree up to order $q^{41}$ (first 12 terms).  While it is impossible to prove
or disprove that the thresholds for the fcc and hcp lattices are identical from
these results, they clearly 
suggest that if the thresholds are indeed different, then
they should be much closer for bond percolation than site percolation,
as indeed we have found numerically.

\section{Conclusions}
\label{sec4}

As a result of this work, we have shown that the critical thresholds for site percolation
on the two lattices are definitely different.  The value of
$p_c$ on the hcp lattice ($0.199\,255\,5 \pm 0.000\,001\,0$) is nearly ten combined error bars away
from the previously reported value \cite{cdlrzjpa} for the fcc lattice ($0.199\,236\,5 \pm
0.000\,001\,0$).  Also, the exact enumeration series for the two lattices begin to
differ at relatively low order terms for site percolation.

On the other hand, even at high precision, the critical thresholds for bond percolation on the hcp
and fcc lattices have the same value, within the error bars ($0.120\,164\,0 \pm
0.000\,001\,0$).  Although the series for bond percolation on the two lattices do differ, the
difference is incredibly small and does not occur until higher order terms.  While this difference
does not rigorously rule out equaltiy of the thresholds, we guess that the thresholds are in fact
slightly different, but by an amount too small to be seen in our numerical simulations.

\end{document}